\documentclass[prd,superscriptaddress,onecolumn,nofootinbib]{revtex4-2}
\usepackage{color}
\usepackage{bm}
\usepackage{graphicx}  
\usepackage{dcolumn}   
\usepackage{bm}        
\usepackage[english]{babel}
\usepackage{booktabs,multirow,array}
\usepackage[bottom]{footmisc}
\usepackage{dblfnote}
\usepackage{mathtools}
\usepackage{amsmath}
\usepackage{physics}
\DFNalwaysdouble 
\usepackage[colorlinks]{hyperref}
\definecolor{LinkColor}{rgb}{0.75, 0, 0}
\definecolor{CiteColor}{rgb}{0, 0.5, 0.5}
\definecolor{UrlColor}{rgb}{0, 0, 0.75}
\hypersetup{linkcolor=LinkColor}
\hypersetup{citecolor=CiteColor}
\hypersetup{urlcolor=UrlColor}
\begin{document}

\title{Hawking radiation in multi-horizon spacetimes using Hamilton Jacobi method}
\author{Chiranjeeb Singha}
\email{chiranjeeb.singha@saha.ac.in}
\affiliation{Theory Division, Saha Institute of Nuclear Physics. 1/AF Bidhan Nagar,
Kolkata 700064, India}

\author{Pritam Nanda}
\email{pritam.nanda@saha.ac.in}
\affiliation{Theory Division, Saha Institute of Nuclear Physics. 1/AF Bidhan Nagar,
Kolkata 700064, India}
\affiliation{ Homi Bhabha National Institute,
Training School Complex, Anushaktinagar, Mumbai 400094, India}

\author{Pabitra Tripathy}
\email{pabitra.tripathy@saha.ac.in}
\affiliation{Theory Division, Saha Institute of Nuclear Physics. 1/AF Bidhan Nagar,
Kolkata 700064, India}
\affiliation{ Homi Bhabha National Institute,
Training School Complex, Anushaktinagar, Mumbai 400094, India}



\begin{abstract}

It has been recently shown that the contribution between the horizons determines the Hawking temperature for a multi-horizon spacetime.
In this article, we apply the Hamiltonian Jacobi method to compute the Hawking temperature for some multi-horizon spacetimes like Schwarzschild-de Sitter spacetime (SdS), Reissner-Nordstrom-de Sitter spacetime (RNdS), and rotating BTZ black hole spacetime (RBTZ) and also arrive at the same conclusion. There are two contributions to the tunneling process of radiation. The combination of these two contributions gives the radiation with the Hawking temperature with an effective surface gravity.

 \end{abstract}
\maketitle
\section{introduction}

There is plenty amount of work has been done to study Hawking radiation as quantum tunneling \cite{Parikh:1999mf,Volovik:2021upi,Volovik:2021iim,Akhmedov:2006pg, PhysRevD.60.024007, Chatterjee:2007hc}. As a most common approach, one can write the tunneling probability as $\Gamma\propto exp~(2\Im[\int p dr])$. But one can argue that this tunneling probability is not a proper observable since the quantity $\int p dr$ is not remaining invariant under canonical transformation. To remove this issue, one can use $\Gamma\propto exp ~(\Im[\oint p dr])$, where the exponential factor is a clearly canonical invariant quantity. If one proceeds with the latter, she may find out the double Hawking temperature as a result, even in the Panilev´e frame (which is regular at the horizon). To overcome this ambiguity, one can use the Hamilton-Jacobi equation method to study Hawking radiation as quantum tunneling. Recently thermodynamics of multi-horizon spacetimes have been studied by one of the authors where $\Gamma\propto exp~(2\Im[\int p dr])$ is used to calculate Hawking temperature for Schwarzschild-de Sitter spacetime, Reissner-Nordstrom-de Sitter spacetime, and rotating BTZ black hole\cite{Singha:2021dxe}. In this paper, we re-investigated the same work with the Hamilton-Jacobi approach, which is free from the canonical invariant issue. In both approaches, the results agree very well.

 It is shown that the contribution between the horizons determines the Hawking temperature for a multi-horizon spacetime \cite{Volovik:2021upi, Volovik:2021iim, Choudhury:2004ph, Chabab:2020xwr, Shankaranarayanan:2003ya, Singha:2021dxe,Azarnia:2021vhc}. It has also been shown that there are two contributions to the tunneling process of radiation for Schwarzschild black hole and de-Sitter spacetime using the Hamilton Jacobi equation method \cite{Volovik:2022vvi}. 
 The combination of these two contributons gives the radiation with the exact Hawking temperature.
Here, we consider the SdS, RNdS, and RBTZ. We compute the Hawking temperature for these spacetimes using the Hamilton-Jacobi equation method. There are two contributions to the tunneling process of radiation. 
The combination of these two contributions gives the radiation with the Hawking temperature with an effective surface gravity. We show that the temperature does not match with the conventional Hawking temperature related to the outer horizon. The contribution between the horizons determines
the Hawking temperature.

In Sec. \ref{Schwarzschild-de Sitter}, we compute the Hawking radiation from the SdS using the Hamilton Jacobi equation \cite{Akhmedov:2006pg, PhysRevD.60.024007, Chatterjee:2007hc}. There are two contributions to the tunneling process of radiation. The combination of these two contributions gives the radiation with the Hawking temperature with an effective surface gravity. We show that the temperature of radiation does not match with the conventional Hawking temperature related to the outer cosmological horizon. The contribution between the two horizons of this spacetime determines the Hawking temperature.

In Sec. \ref{Reissner-Nordstrom-de Sitter},  we compute the Hawking radiation from the RNdS using the Hamilton Jacobi equation \cite{Akhmedov:2006pg, PhysRevD.60.024007, Chatterjee:2007hc}. There are two contributions to the tunneling process of radiation. The combination of these two contributions gives the radiation with the Hawking temperature with an effective surface gravity. We show that the temperature of radiation does not match with the conventional Hawking temperature related to the outer cosmological horizon. The contribution between the three horizons of this spacetime determines the Hawking temperature.

In Sec. \ref{BTZ}, we compute the Hawking radiation from the RBTZ using the Hamilton Jacobi equation \cite{Akhmedov:2006pg, PhysRevD.60.024007, Chatterjee:2007hc}. There are two contributions to the tunneling process of radiation. The combination of these two contributions gives the radiation with the Hawking temperature with an effective surface gravity. We show that the temperature of radiation does not match with the conventional Hawking temperature related to the outer event horizon. The contribution between the two horizons of this spacetime determines
the Hawking temperature. 

\section{Hawking radiation in SdS}\label{Schwarzschild-de Sitter}

The metric for the SdS is given by \cite{Bhattacharya:2013tq, Shankaranarayanan:2003ya, Medved:2002zj, Pappas:2017kam, Robson:2019yzx,Tian:2003ua, Singha:2021dxe},
\begin{equation}\label{eq1}
ds^2=-f(r)dt^2+f(r)^{-1} dr^2 +r^2 d \Omega^2~,
\end{equation}
where $f(r)=\left(1-\frac{2 M}{r}-\frac{r^2}{l^2}\right)$. Here $M$ is the mass of the black hole and $l^2= \frac{3}{\Lambda}$, where $\Lambda$ is the positive cosmological constant. One can show that there are two horizons present in this spacetime if the condition $0<x<1/27$ holds true. Here $x=M^2/l^2$. These two horizons are the black hole horizon ($r_{H}$) and the cosmological horizon ($r_{C}$), respectively. The inner horizon is $r_{H}$, and the outer horizon is $r_{C}$ in this spacetime. The expression for $r_{H}$ and $r_{C}$ are given by Appendix \ref{appendix1},
\begin{align}
r_{H}=\frac{2 M}{\sqrt{3 x}} \cos \frac{\pi+\phi}{3}~,\\
 r_{C}=\frac{2 M}{\sqrt{3 x}} \cos \frac{\pi-\phi}{3}~,
\end{align}
where $\phi= \cos^{-1}(3 \sqrt{3x})$. The surface gravities at the black hole horizon ($\kappa_{H}$) and the cosmological horizon ($\kappa_{C}$) are given by \cite{Bousso:1997wi, Shankaranarayanan:2003ya}, 
\begin{align}
\kappa_{H}= a \left|\frac{M}{r^2_H}-\frac{r_H}{l^2}\right|~,\\
\kappa_{C}= a \left|\frac{M}{r^2_C}-\frac{r_C}{l^2}\right|~.
\end{align}
Here $a = 1/\sqrt{1-(27x)^{1/3}}$. If one transforms the coordinate as,
\begin{equation}\label{eq6}
 d\tilde{t}\rightarrow dt \pm f dr,~~f=\frac{\sqrt{\frac{2M}{r}+\frac{r^2}{l^2}}}{\left(1-\frac{2 M}{r}-\frac{r^2}{l^2}\right)}~.
\end{equation}
Then the Painleve-Gullstrand (PG) metric for SdS is obtained. The PG metric for Sds is given by \cite{Singha:2021dxe},
\begin{equation}\label{eq7}
ds^2= g_{\mu \nu} dx^{\mu} dx^{\nu}=-dt^2+(dr\pm v dt)^2+r^2 d\Omega^2~.
\end{equation}
Here $v$ is the shift velocity. The expression of $v^2$ is given by,
\begin{equation}\label{eqn8}
 v^2=\frac{2 M}{r}+\frac{r^2}{l^2}~.
\end{equation}

Now we consider a massive particle. The field equation for the massive particle with mass $m$ in a tunneling trajectory is given by,
\begin{equation}
 \frac{-\hbar^2}{\sqrt{-g}}\partial_{\mu}\bigg(\sqrt{-g}g^{\mu\nu}\partial_\nu\bigg)\phi + m^2\phi=0~.  
\end{equation}
We also consider $\phi=e^{\frac{i}{\hbar}S}$ which leads to Hamilton Jacobi equation as,
\begin{equation}
    -i \hbar\bigg(\frac{1}{\sqrt{-g}}\partial_{\mu}\big(\sqrt{-g}g^{\mu\nu}\big)\partial_\nu S+ g^{\mu\nu}\partial_\mu\partial_\nu S\bigg)+g^{\mu\nu}\partial_\mu S\partial_\nu S+m^2=0~.
\end{equation}
Now taking the limit $\hbar\rightarrow0$ and compering $O(1)$ term we get
\begin{equation}
 g^{\mu\nu}\partial_\mu S\partial_\nu S + m^2=0~.
\end{equation}

Here $g^{\mu \nu}$ is the contravariant metric, there is a killing field $t^a=(1,0,0,0)$ with respect to which we define energy $E$. Formulation of hamilton jacobi equation allows us to write $S=Et+S_r(r)$. Now we can expand the above equation in PG coordinate as

\begin{equation}
-E^2+(1-v^2)\bigg(\frac{dS_r}{dr}\bigg)^2+2vE\frac{dS_r}{dr}+m^2=0~.
\end{equation}
We write then the solution of the above differential equation as an integral form which is given by,
\begin{equation}\label{eqn12}
    S_r=-\int\frac{Ev}{1-v^2}dr\pm\int\frac{\sqrt{E^2-m^2(1-v^2)}}{1-v^2}dr~. 
\end{equation}
Now putting the expression for $v$ (\ref{eqn8}) in Eq. (\ref{eqn12}) we get
\begin{equation} \label{eq:14}
    S_r=-\int\frac{E\sqrt{\frac{2 M}{r}+\frac{r^2}{l^2}}}{1-\frac{2 M}{r}-\frac{r^2}{l^2}}dr\pm\int\frac{\sqrt{E^2-m^2(1-\frac{2 M}{r}-\frac{r^2}{l^2})}}{1-\frac{2 M}{r}-\frac{r^2}{l^2}}dr~. 
\end{equation}
In Eq. $(\ref{eq:14})$,  the $\pm$ sign corresponds to the fact that it includes both the incoming and outgoing solution.

The expression $\left(1-\frac{2 M}{r}-\frac{r^2}{l^2}\right)^{-1}$ can also be expressed as ( for more details, see Appendix \ref{appendix2} ),
\begin{eqnarray}
\left(1-\frac{2 M}{r}-\frac{r^2}{l^2}\right)^{-1}=\frac{a}{2 \kappa_{H}(r-r_{H})}+\frac{a}{2 \kappa_{C}(r-r_{C})}-\frac{a}{2 \kappa_{0}(r-r_{0})}~.\nonumber\\
\end{eqnarray}
Here $r_{0}=-(r_{C}+r_{H})$ is a unphysical horizon and surface gravity associated with $r_{0}$ is defined as $\kappa_0=\frac{1}{2}\left|\frac{\partial f(r)}{\partial r}\right|_{r=r_0}$. So, in Eq. $(\ref{eq:14})$,  there are three pole at three horizon and we consider only two as third one is unphysical . 
Im $S$ can also be rewritten as,
\begin{eqnarray}
Im~S_r &=& -\int\frac{E\sqrt{\frac{2 M}{r}+\frac{r^2}{l^2}}}{1-\frac{2 M}{r}-\frac{r^2}{l^2}}dr\pm\int\frac{\sqrt{E^2-m^2(1-\frac{2 M}{r}-\frac{r^2}{l^2})}}{1-\frac{2 M}{r}-\frac{r^2}{l^2}}dr \nonumber\\ 
&=& - Im \int E \sqrt{\frac{2 M}{r}+\frac{r^2}{l^2}}\times \left(\frac{a}{2 \kappa_{H}(r-r_{H})}+\frac{a}{2 \kappa_{C}(r-r_{C})} -\frac{a}{2 \kappa_{0}(r-r_{0})}\right)dr\nonumber\\
&\pm & Im \int  \sqrt{E^2-m^2\left(1-\frac{2 M}{r}-\frac{r^2}{l^2}\right)}\times \left(\frac{a}{2 \kappa_{H}(r-r_{H})}+\frac{a}{2 \kappa_{C}(r-r_{C})} -\frac{a}{2 \kappa_{0}(r-r_{0})}\right)dr~.\label{eq12}
\end{eqnarray}
From Eq. (\ref{eq12}), one can easily show that the contribution of two horizons for this spacetime (\ref{eq1}) gives the probability of the Hawking radiation as ( for more details, see Appendix \ref{appendix3} ),
\begin{equation}\label{eq13}
P=e^{\left(- \frac{ \pi a E}{\kappa_{eff}}\right)}\times e^{\left(\pm \frac{ \pi a E}{\kappa_{eff}}\right)}~.
\end{equation}
Here $\kappa_{eff}=\left(\frac{1}{\kappa_{H}}+\frac{1}{\kappa_{C}}\right)^{-1}$. We get a nontrivial probability distribution, for the negative sign of the second exponential in Eq. $(\ref{eq13})$. This corresponds to thermal radiation. The temperature of this thermal radiation is given by \cite{Singha:2021dxe},
\begin{equation}\label{eq14}
T_{H}=\frac{\kappa_{eff}}{2 \pi a}= \frac{\kappa_{C}\kappa_{H}}{2 \pi a (\kappa_{C}+\kappa_{H})}~.
\end{equation}
We show that there are two contributions to the tunneling process of radiation. The combination of these two contributions gives the radiation with the Hawking temperature with an effective surface gravity.
Eq. (\ref{eq14}) also implies that the temperature of radiation does
not match with the conventional Hawking temperature related to the outer cosmological horizon. The contribution
between the two horizons of this spacetime determines the Hawking temperature.
\section{HAWKING RADIATION FROM RNdS}\label{Reissner-Nordstrom-de Sitter}

The metric for the RNdS is given by \cite{Li:2021axp, Zhang:2016nws, Hollands:2019whz, Guo:2005hw, Ahmed:2016lou, Singha:2021dxe},
\begin{equation}\label{eq23}
ds^2=-f(r)dt^2+f(r)^{-1} dr^2 +r^2 d \Omega^2~.
\end{equation}
where $f(r)=\left(1-\frac{2 M}{r}+\frac{q^2}{r^2}-\frac{\Lambda r^2}{3}\right)$. Here $M$ is the mass of the black hole, $q$ is the charge of the black hole, and $\Lambda$ is the cosmological constant which is taken to be positive. There are three horizons for this spacetime (\ref{eq23}). Here we denote the event horizon as $r_{+}$ , Cauchy horizon as $r_{-}$ and the outer  cosmological horizon as $r_{C}$ and the surface gravities at the event horizon ($\kappa_{+}$), the Cauchy horizon ( $\kappa_{-}$) and the cosmological horizon ( $\kappa_{C}$). The expressions of the surface gravities are given by \cite{Singha:2021dxe},
\begin{eqnarray}
\kappa_{+}&=& \left|\frac{M}{r^2_+}- \frac{q^2}{r^3_{+}}-\frac{\Lambda r_+}{3}\right|~,\nonumber\\
\kappa_{-}&=& -\left|\frac{M}{r^2_-}- \frac{q^2}{r^3_{-}}-\frac{\Lambda r_-}{3}\right|~,\nonumber\\
\kappa_{C}&=& \left|\frac{M}{r^2_C}- \frac{q^2}{r^3_{C}}-\frac{\Lambda r_C}{3}\right|~.
\end{eqnarray}
One can get the PG metric for RNdS using the following coordinate transformations,
\begin{equation}\label{eq24}
 d\tilde{t}\rightarrow dt \pm f dr,~~f=\frac{\sqrt{{\left(\frac{2M}{r}-\frac{q^2}{r^2}+\frac{\Lambda r^2}{3}\right)}}}{\left(1-\frac{2 M}{r}+\frac{q^2}{r^2}-\frac{\Lambda r^2}{3}\right)}~,
\end{equation}
where the PG metric for RNdS is given by,
\begin{eqnarray}\label{eq21}
ds^2 = g_{\mu \nu} dx^{\mu} dx^{\nu}=-dt^2+(dr\pm v dt)^2+r^2 d\Omega^2~.
\end{eqnarray}
Here the shift velocity is $v$. The expression of $v^2$ is given by,
\begin{equation}
 v^2={\left(\frac{2M}{r}-\frac{q^2}{r^2}+\frac{\Lambda r^2}{3}\right)}~.
\end{equation}
Now we could perform a similar calculation for the tunneling trajectory for a massive particle in the  RNdS. So we write the solution of Hamilton Jacobi equation as,

\begin{equation}
    S_r=-\int\frac{E\sqrt{{\left(\frac{2M}{r}-\frac{q^2}{r^2}+\frac{\Lambda r^2}{3}\right)}}}{1-\frac{2 M}{r}+\frac{q^2}{r^2}-\frac{\Lambda r^2}{3}}dr
    \pm\int\frac{\sqrt{E^2-m^2(1-\frac{2 M}{r}+\frac{q^2}{r^2}-\frac{\Lambda r^2}{3})}}{1-\frac{2 M}{r}+\frac{q^2}{r^2}-\frac{\Lambda r^2}{3}}dr~. 
\end{equation}

The expression, $\left(1-\frac{2 M}{r}+\frac{q^2}{r^2}-\frac{\Lambda r^2}{3}\right)^{-1}$, can also be expressed as \cite{Singha:2021dxe},
\begin{eqnarray}
\left(1-\frac{2 M}{r}+\frac{q^2}{r^2}-\frac{\Lambda r^2}{3}\right)^{-1}=\frac{1}{2 \kappa_{+}(r-r_{+})}+\frac{1}{2 \kappa_{-}(r-r_{-})} +\frac{1}{2 \kappa_{C}(r-r_{C})}-\frac{1}{2 \kappa_{0}(r-r_{0})}~.
\end{eqnarray}
Here $r_{0}=-(r_{+}+r_{-}+r_{C})$ is unphysical horizon and surface gravity associated with $r_{0}$ is defined as $\kappa_0=\frac{1}{2}\left|\frac{\partial f(r)}{\partial r}\right|_{r=r_0}$.
So, Im $S$ can also be rewritten as,
\begin{eqnarray}
Im~S_r &=& Im \Bigg[-\int\frac{E\sqrt{{\left(\frac{2M}{r}-\frac{q^2}{r^2}+\frac{\Lambda r^2}{3}\right)}}}{1-\frac{2 M}{r}+\frac{q^2}{r^2}-\frac{\Lambda r^2}{3}}dr
\pm\int\frac{\sqrt{E^2-m^2(1-\frac{2 M}{r}+\frac{q^2}{r^2}-\frac{\Lambda r^2}{3})}}{1-\frac{2 M}{r}+\frac{q^2}{r^2}-\frac{\Lambda r^2}{3}}dr\Bigg]\nonumber\\ 
&=& - Im \int E\sqrt{{\left(\frac{2M}{r}-\frac{q^2}{r^2}+\frac{\Lambda r^2}{3}\right)}}\times  \left(\frac{1}{2 \kappa_{+}(r-r_{+})}+\frac{1}{2 \kappa_{-}(r-r_{-})}+\frac{1}{2 \kappa_{C}(r-r_{C})}-\frac{1}{2 \kappa_{0}(r-r_{0})}\right)dr~\nonumber\\
&&\pm Im \int \sqrt{E^2-m^2\bigg(1-\frac{2 M}{r}+\frac{q^2}{r^2}-\frac{\Lambda r^2}{3}\bigg)}\times   \left(\frac{1}{2 \kappa_{+}(r-r_{+})}+\frac{1}{2 \kappa_{-}(r-r_{-})}+\frac{1}{2 \kappa_{C}(r-r_{C})}-\frac{1}{2 \kappa_{0}(r-r_{0})}\right)dr~.\nonumber\\\label{eq33}
\end{eqnarray}
From Eq. (\ref{eq33}), one can easily show that the contribution of three horizons gives the probability of the Hawking radiation as,
\begin{equation}\label{eq32}
P=e^{\left(- \frac{ \pi E}{\kappa_{eff}}\right)}\times e^{\left(\pm \frac{ \pi E}{\kappa_{eff}}\right)},
\end{equation}
where $\kappa_{eff}=\left(\frac{1}{\kappa_{+}}+\frac{1}{\kappa_{-}}+\frac{1}{\kappa_{C}}\right)^{-1}$. If we follow the outgoing trajectory we get nontrivial distribution from Eq. (\ref{eq32}). This corresponds to the thermal radiation. The temperature of this thermal radiation is given by \cite{Singha:2021dxe},
\begin{equation}\label{eq31}
T_{H}=\frac{\kappa_{eff}}{2 \pi}~.
\end{equation}
Here we show that there are two contributions to the tunneling process of radiation. The combination of these two contributions gives the
radiation with the Hawking temperature with an effective surface gravity.
Eq. (\ref{eq31}) also implies that the temperature of radiation does not
match with the conventional Hawking temperature related to the outer cosmological horizon. The contribution
between the three horizons of this spacetime determines the Hawking temperature.
 \section{HAWKING RADIATION in RBTZ}\label{BTZ}

The metric for a RBTZ is given by \cite{Banados:1992wn,Banados:1992gq,Dias:2019ery, Chaturvedi:2013ova, Kajuri:2020bvi,Fathi:2021eig, Bhattacharjee:2020gbo, Emparan:2020rnp, Singha:2021dxe},
\begin{equation}\label{eq45}
ds^2=-f(r)dt^2+f(r)^{-1} dr^2 +r^2 \left(d \phi-\frac{J}{2r^2}dt\right)^2~,
\end{equation}
where $f(r)=\left(-M+\frac{r^2}{l^2}+\frac{J^2}{4 r^2}\right)=\frac{(r^2-r^2_{+})(r^2-r^2_{-})}{l^2 r^2}$.  Here $M$ is the mass of the RBTZ black hole, $\Lambda=-\frac{1}{l^2}$, where $\Lambda$ is the negative cosmological constant, $J$ is the angular momentum and $r_{\pm}=l~\left(\frac{M}{2}\left(1\pm\sqrt{1-(\frac{J}{M l})^2}\right)\right)^{1/2}$ is the inner Cauchy horizon and the outer event horizon, respectively. To get the PG metric for RBTZ first we move to a dragging coordinate system with an angular velocity as \cite{Liu:2005hj},
\begin{equation}
    \frac{d\phi}{dt}=\frac{J}{2r^2}~.
\end{equation} 
In this coordinate system the line element becomes
\begin{equation}
    ds^2=-f(r)dt^2+f(r)^{-1} dr^2~.
\end{equation} 
which represent two dimensional hypersurface in a three dimensional BTZ space time. Now using following coordinate transformations,
\begin{equation}\label{eq46}
 d\tilde{t}\rightarrow dt \pm f dr,~~f=\frac{\sqrt{1+M-\frac{r^2}{l^2}-\frac{J^2}{4 r^2}}}{\left(-M+\frac{r^2}{l^2}+\frac{J^2}{4 r^2}\right)}~,
\end{equation}
we obtain the PG metric for RBTZ as,
\begin{eqnarray}\label{eq47}
ds^2= g_{\mu \nu} dx^{\mu} dx^{\nu}=-dt^2+(dr\pm v dt)^2+r^2 d\Omega^2~. \nonumber\\
\end{eqnarray}
Here the shift velocity is $v$ and the expression of $v^2$ is given by,
\begin{equation}
 v^2=1+M -\frac{r^2}{l^2}-\frac{J^2}{4 r^2}~.
\end{equation}

The exponent of the imaginary part of the action along the tunneling trajectory, Im $S$, gives the probability of the tunneling process, Where action, $S$, is the solution of Hamilton-Jacobi equation which is given by,
\begin{equation}
\begin{split}
    S_r=-\int\frac{E\sqrt{1+M-\frac{r^2}{l^2}-\frac{J^2}{4 r^2}}}{-M+\frac{r^2}{l^2}+\frac{J^2}{4 r^2}}dr\pm\int\frac{\sqrt{E^2-m^2(-M+\frac{r^2}{l^2}+\frac{J^2}{4 r^2})}}{-M+\frac{r^2}{l^2}+\frac{J^2}{4 r^2}}dr~. 
    \end{split}
\end{equation}
As  the expression, $\left(-M+\frac{r^2}{l^2}+\frac{J^2}{4 r^2}\right)^{-1}$, can be expressed as \cite{Singha:2021dxe},
\begin{eqnarray}
\left(-M+\frac{r^2}{l^2}+\frac{J^2}{4 r^2}\right)^{-1}=\frac{l^2 r^2}{(r^2-r^2_{+})(r^2-r^2_{-})}~.\nonumber\\
\end{eqnarray}
So, Im $S$ can also be rewritten as,
\begin{eqnarray}
Im~S_r &=& Im \Bigg[-\int\frac{E\sqrt{1+M-\frac{r^2}{l^2}-\frac{J^2}{4 r^2}}}{-M+\frac{r^2}{l^2}+\frac{J^2}{4 r^2}} dr\pm \int\frac{\sqrt{E^2-m^2(-M+\frac{r^2}{l^2}+\frac{J^2}{4 r^2})}}{-M+\frac{r^2}{l^2}+\frac{J^2}{4 r^2}}dr\Bigg] \nonumber\\&&= - Im \int E\sqrt{1+M-\frac{r^2}{l^2}-\frac{J^2}{4 r^2}}\times\frac{l^2 r^2}{(r^2-r^2_{+})(r^2-r^2_{-})}dr\nonumber\\
&&\pm  Im \int \sqrt{E^2-m^2\left(-M+\frac{r^2}{l^2}+\frac{J^2}{4 r^2}\right)}\times\frac{l^2 r^2}{(r^2-r^2_{+})(r^2-r^2_{-})}dr~. \label{eq52}
\end{eqnarray}
From Eq. (\ref{eq52}), one can easily show that the contribution of two horizons gives the probability of Hawking radiation as,
\begin{equation}\label{eq50}
P=e^{\left(- \frac{ \pi l^2~E}{(r_{+}+r_{-})}\right)}\times e^{\left(\pm \frac{ \pi l^2~E}{(r_{+}+r_{-})}\right)}~.
\end{equation}
 Similarly here also we only consider the negative sign in equation (\ref{eq50}) to get a nontrivial distribution. This corresponds to thermal radiation. The temperature of the thermal radiation is given by \cite{Singha:2021dxe},
\begin{equation}\label{eq54}
T_{H}=\frac{(r_{+}+r_{-})}{2 \pi l^2}~.
\end{equation}
This temperature can also expressed as \cite{Singha:2021dxe},
\begin{equation}
T_{H}=\frac{\kappa_{eff}}{2 \pi}~,   
\end{equation}
where $\kappa_{eff}=\left(\frac{1}{\kappa_{+}}+\frac{1}{\kappa_{-}}\right)^{-1}$. Here we define $\kappa_{+}$ and $\kappa_{-}$ as the surface gravities of the event horizon and the Cauchy horizon. 
We show that there are two contributions to the tunneling process of radiation. The combination of these two contributions gives the
radiation with the Hawking temperature with an effective surface gravity. 
Eq. (\ref{eq54}) also implies that the temperature of radiation that does not match with
the conventional Hawking temperature related to the outer event horizon. The contribution
between the two horizons of this spacetime determines the Hawking temperature.

\section{DISCUSSIONS}

The most common approach to studying Hawking radiation in quantum tunneling is to write the tunneling probability as $\Gamma\propto exp(2\Im[\int p dr])$. But this tunneling probability is not an observable since the quantity $\int p dr$ is not remaining invariant under canonical transformation. To remove this issue, one can use $\Gamma\propto exp(\Im[\oint p dr])$, where the exponential factor is a clearly canonical invariant quantity. Here we use the Hamilton-Jacobi equation method, which is free from the canonical invariant issue, to study Hawking radiation as tunneling. Here we have considered SdS, RNdS, and RBTZ. We have shown that there are two contributions to the tunneling process of radiation. The combination of these two contributions gives the radiation with the Hawking temperature with an effective surface gravity. We have also shown that the temperature of radiation
 does not match with the conventional Hawking temperature related to the outer cosmological horizon for SdS. The contribution
between the two horizons of this spacetime determines the Hawking temperature.
 For RNdS, the temperature of radiation does not match with the conventional Hawking temperature related to the outer cosmological horizon. The contribution
between the three horizons of this spacetime determines the Hawking temperature. 
Similarly, for RBTZ,   the temperature of radiation does
not match with the conventional Hawking temperature related to the outer event horizon. The contribution
between the two horizons of this spacetime determines the Hawking temperature.

It would be interesting to apply the above technique for other spacetimes \cite{Boyer:1966qh, Kerr:2007dk, Krasinski:1976vyc, Teukolsky:2014vca, Visser:2007fj, Smailagic:2010nv, Akcay:2010vt,Li:2016zdi,Suzuki:1998vy, Franzen:2020gke,Gwak:2018tmy,Stuchlik:1997gk} for defining global temperature for those spacetimes. We leave this for the future.

One can also try to calculate the Hawking temperature with different modifications of spacetime and matter field \cite{ARAUJOFILHO2023137744, 2019EPJC...79..358H, HEIDARI2023137707, 10.1093/ptep/ptz085, 2022EPJP..137.1147J, CHEN2022136994, MAGHSOODI2020100559, article, 2020EPJC...80..696F} using the above technique. It will be nice to check whether can a global temperature exists or not for those spacetimes.

\begin{acknowledgments}

 CS thanks the Saha Institute of Nuclear Physics (SINP) Kolkata for financial support.
 
\end{acknowledgments}

\appendix
\section{Calculation for finding the horizon radii $r_H$ and $r_C$}\label{appendix1}

We start with by conserding the following equation
\begin{equation}
\begin{split}
    &1-\frac{2M}{r}-\frac{r^2}{l^2}=0\\
    &r^3-rl^2+2Ml^2=0~.
    \end{split}
\end{equation}
To get three real roots of this cubic equation the discriminant should be positive. The condition can be expressed as,
\begin{equation}
\begin{split}
   &-(4(-l^2)^3+27(2Ml^2)^2)>0\\
   &0<\frac{M^2}{l^2}<\frac{1}{27}~.\\
\end{split}
\end{equation}
Now we choose $r=\alpha\cos\theta$
\begin{equation}
\begin{split}
    &\alpha^3\cos^3\theta-l^2\alpha\cos\theta+2Ml^2=0\\
    &\alpha^3\frac{\cos3\theta+3\cos\theta}{4}-l^2\alpha\cos\theta+2Ml^2=0\\
&\frac{\alpha^3}{4}\cos3\theta+\bigg(\frac{3\alpha^3}{4}-l^2\alpha\bigg)\cos\theta+2Ml^2=0~.
    \end{split}
\end{equation}
We can choose $\alpha$ in such a way that the second term becomes zero. Here we consider $\alpha$ as,
\begin{equation}
    \alpha=\frac{2}{\sqrt{3}}l~.
\end{equation}
From the last line of equation $(2)$, we get
\begin{eqnarray}
\frac{\alpha^3}{4}\cos3\theta+2Ml^2 &=&0 \nonumber\\
     \cos3\theta &=& -\frac{8M}{\alpha^3}l^2\nonumber\\
    \cos(2\pi k +3\theta)&=& -\frac{8M}{\alpha^3}l^2~,
\end{eqnarray}    
where
\begin{eqnarray} 
 k \in Z; \;\;
    \theta =\frac{\pi-2\pi k-\cos^{-1}\bigg(\frac{8M}{\alpha^3}l^2\bigg)}{3}~.
\end{eqnarray}
Define $\phi=\cos^{-1}(3\sqrt{3x})$ and $x=\frac{M^2}{l^2}$, the above equation can be written as, 
\begin{equation}
    \theta=\frac{\pi-2\pi k-\phi}{3}~.
\end{equation}
 Three real roots  of equation $(1)$ corresponds to $k=0, k=1 \;\text{and} \;k=-1$ are respectively,
\begin{equation}
    r_{C}=\frac{2M}{\sqrt{x}}\cos\bigg(\frac{\pi-\phi}{3}\bigg)~,
\end{equation}
\begin{equation}
    r_{H}=\frac{2M}{\sqrt{x}}\cos\bigg(\frac{\pi+\phi}{3}\bigg)~,
\end{equation}
and
\begin{equation}
    r_{0}=-\frac{2M}{\sqrt{x}}\cos\bigg(\frac{\phi}{3}\bigg)~.
\end{equation}
We discard the last solution as the horizon can not be at a negative value of radial coordinate.

 \section{Derivation of equation $(15)$ }\label{appendix2}

 Equation $(15)$ can be written as,
 \begin{equation}
    \bigg(1-\frac{2M}{r}-\frac{r^2}{l^2}\bigg)^{-1}=\frac{rl^2}{l^2r-2Ml^2-r^3}~.
\end{equation}
Lets $r_H,r_C$, and $r_0$ be the roots of the equation $l^2r-2Ml^2-r^3=0$. We can express r as 
\begin{equation}
    r=A(r-r_H)(r-r_c)+B(r-r_H)(r-r_0)+C(r-r_c)(r-r_0)~.
\end{equation}
Now choosing $r=r_0$, $r_C$ ,  and $r_H$, we get
\begin{equation}
    A=\frac{r_0}{(r_0-r_H)(r_0-r_c)}~,
\end{equation}

\begin{equation}
    B=\frac{r_C}{(r_C-r_H)(r_C-r_0)}~,
\end{equation}
and
\begin{equation}
    C=\frac{r_H}{(r_H-r_C)(r_H-r_0)}~.
\end{equation}
Thus, equation $(15)$ can be written as,
\begin{equation}
  \begin{split}
     \bigg(1-\frac{2M}{r}-\frac{r^2}{l^2}\bigg)^{-1}&=\frac{A(r-r_H)(r-r_c)+B(r-r_H)(r-r_0)+C(r-r_c)(r-r_0)}{(r-r_H)(r-r_C)(r-r_0)}l^2\\
     &=\frac{A l^2}{r-r_0}+\frac{B l^2}{r-r_C}+\frac{C l^2}{r-r_H}~.
     \end{split}
\end{equation}
We have 
\begin{equation}
\begin{split}
f(r)&=\bigg(1-\frac{2M}{r}-\frac{r^2}{l^2}\bigg)\\
&=\frac{rl^2-2Ml^2-r^3}{rl^2}\\
&=\frac{(r-r_H)(r-r_C)(r-r_0)}{rl^2}~.
\end{split}
\end{equation}
Taking derivative of equation $(16)$ we get,
\begin{equation}
    f'(r)=-\frac{(r-r_H)(r-r_C)(r-r_0)}{r^2l^2}+\frac{(r-r_C)(r-r_0)}{rl^2}+ \frac{(r-r_H)(r-r_0)}{rl^2}+\frac{(r-r_H)(r-r_C)}{rl^2}~.
\end{equation}

Surface gravity associated with $r_{0}$ is defined as $\kappa_0=\frac{1}{2}\left|\frac{\partial f(r)}{\partial r}\right|_{r=r_0}$.
Using equation(17) we find the expression of $\kappa_0$,
\begin{equation}
\begin{split}
 \kappa_0=&\frac{1}{2}f'(r_0)\\
 =&\frac{(r_0-r_H)(r_0-r_C)}{2r_0l^2}\\
 =&\frac{1}{2Al^2}~.
\end{split}    
\end{equation}
Similarly surface gravity terms at $r=r_C$ and $r=r_H$ are respectively,
\begin{equation}
    \kappa_C=\frac{1}{2Bl^2}~,
\end{equation}
and 
\begin{equation}
    \kappa_H=\frac{1}{2Cl^2}~.
\end{equation}
Now we substitute A,B and C from equations $(18)$, $(19)$, and $(20)$ into equation $(15)$. We get, 
\begin{equation}
    \bigg(1-\frac{2M}{r}-\frac{r^2}{l^2}\bigg)^{-1}=\frac{1}{2\kappa_0(r-r_0)}+\frac{1}{2\kappa_C(r-r_c)}+\frac{1}{2\kappa_H(r-r_H)}~.
\end{equation}

We have also introduced $a = 1/\sqrt{1-(27x)^{1/3}}$ in the SDS case which is a normalised constant. We see that the final Hawking temperature does not depend on the parameter $a$.

Equations (25) and (36) can be derived in the same way.

\section{Calculation of probability (P) for the Hawking radiation.}\label{appendix3}

We start with the following equation,
\begin{equation}
\begin{split}
  ImS = &- Im \int E \sqrt{\frac{2 M}{r}+\frac{r^2}{l^2}}\times \left(\frac{a}{2 \kappa_{H}(r-r_{H})}+\frac{a}{2 \kappa_{C}(r-r_{C})} -\frac{a}{2 \kappa_{0}(r-r_{0})}\right)dr\\
&\pm Im \int  \sqrt{E^2-m^2\left(1-\frac{2 M}{r}-\frac{r^2}{l^2}\right)}\times \left(\frac{a}{2 \kappa_{H}(r-r_{H})}+\frac{a}{2 \kappa_{C}(r-r_{C})} -\frac{a}{2 \kappa_{0}(r-r_{0})}\right)dr~.
\end{split}
\end{equation}
For computational simplicity let's break down the integration.
Let's take,
\begin{equation}
    I_1=\int E \sqrt{\frac{2 M}{r}+\frac{r^2}{l^2}}\times \left(\frac{a}{2 \kappa_{H}(r-r_{H})}\right)dr~.
\end{equation}
It is noted that there is a simple pole at $r=r_H$. Let's take $r-r_H=\epsilon e^{i\theta}$.
\begin{equation*}
\begin{split}
     I_1=&\lim_{\epsilon\to 0}\int E \sqrt{\frac{2M}{r_H+\epsilon e^{i\theta}}+\frac{(r_H+\epsilon e^{i\theta})^2}{l^2}}\times \frac{a}{2\kappa_H \epsilon e^{i\theta}}\times i\epsilon e^{i\theta} d\theta\\
     =&i\frac{Ea\pi}{2\kappa_H}~.
\end{split}
\end{equation*}
Similarly, 
\begin{equation}
    I_2=\int E \sqrt{\frac{2 M}{r}+\frac{r^2}{l^2}}\times \left(\frac{a}{2 \kappa_{C}(r-r_{C})}\right)dr=i\frac{Ea\pi}{2\kappa_C}~.
\end{equation}
Let's take,
\begin{equation}
 I_3= \int  \sqrt{E^2-m^2\left(1-\frac{2 M}{r}-\frac{r^2}{l^2}\right)}\times \left(\frac{a}{2 \kappa_{H}(r-r_{H})}\right) dr~.
\end{equation}
 After change of variable from $r$ to $\theta$ we get,
 \begin{equation}
     \begin{split}
  I_3=&\lim_{\epsilon\to 0} \int  \sqrt{E^2-m^2\left(1-\frac{2 M}{r_H+\epsilon e^{i\theta}}-\frac{(r_H+\epsilon e^{i\theta})^2}{l^2}\right)}\times \frac{a}{2\kappa_H \epsilon e^{i\theta}}\times i\epsilon e^{i\theta} d\theta \\
  =&i\frac{Ea\pi}{2\kappa_H}~. 
     \end{split}
 \end{equation}
Similarly,
\begin{equation}
 I_4= \int  \sqrt{E^2-m^2\left(1-\frac{2 M}{r}-\frac{r^2}{l^2}\right)}\times \left(\frac{a}{2 \kappa_{C}(r-r_{C})}\right) dr=i\frac{Ea\pi}{2\kappa_C}~. 
\end{equation}
We do not keep the third and sixth terms of equation $(22)$ as they are unphysical.
Substituting  values of $I_1,I_2,I_3$ and $I_4$ into equation $(22)$,
\begin{equation}
\begin{split}
     ImS=&-Im\left(i\frac{Ea\pi}{2\kappa_H}+i\frac{Ea\pi}{2\kappa_C}\right)\pm Im\left(i\frac{Ea\pi}{2\kappa_H}+i\frac{Ea\pi}{2\kappa_C}\right)\\
     =&-\frac{Ea\pi}{2}\left(\frac{1}{\kappa_H}+\frac{1}{\kappa_C}\right)\pm \frac{Ea\pi}{2}\left(\frac{1}{\kappa_H}+\frac{1}{\kappa_C}\right)\\
     =&-\frac{Ea\pi}{2\kappa_{eff}}\pm \frac{Ea\pi}{2\kappa_{eff}}~,
    \end{split}
\end{equation}
where, 
\begin{equation*}
    \kappa_{eff}=\left(\frac{1}{\kappa_H}+\frac{1}{\kappa_C}\right)^{-1}~.
\end{equation*}
The probability is given by, 
\begin{equation}
    P\propto exp\left(2ImS\right)=e^{\left(- \frac{ \pi a E}{\kappa_{eff}}\right)}\times e^{\left(\pm \frac{ \pi a E}{\kappa_{eff}}\right)}~.
\end{equation}

Similarly we can derive equation $(27)$ from $(26)$ and equation $(38)$ from $(37)$.

\providecommand{\noopsort}[1]{}\providecommand{\singleletter}[1]{#1}%

\end{document}